\documentclass[conference]{IEEEtran}
\IEEEoverridecommandlockouts

\usepackage{cite}
\usepackage{amsmath,amssymb,amsfonts}
\usepackage{graphicx}
\usepackage{textcomp}
\usepackage{xcolor}
\usepackage{booktabs}
\usepackage{multirow}
\usepackage{tabularx}
\usepackage{url}
\usepackage{hyperref}
\usepackage{balance}
\usepackage[utf8]{inputenc}
\usepackage[whole]{bxcjkjatype}

\newcommand{\model}{\texttt{Llama3-nikkei-genrec}}
\newcommand{\dedupmodel}{\texttt{Llama3-nikkei-genrec-dedup}}
\newcommand{\articlesep}{\texttt{[ARTICLE\_SEP]}}
\newcommand{\tightfigurecaption}{\vspace{-0.30em}}

\begin{document}

\title{Generative News Recommendation Reproduces Popularity: A Memorization-Based Analysis}

\author{\IEEEauthorblockN{Shotaro Ishihara\thanks{This work was conducted while the author was with Nikkei Inc.}}
\IEEEauthorblockA{\textit{Nikkei Inc.}\\
Tokyo, Japan\\
upura0@gmail.com}}

\maketitle

\begin{abstract}
Text-based generative recommenders can go beyond ranking existing items by directly generating item titles or other textual representations. In the news domain, this raises a societal concern that models may repeatedly surface already popular articles and thereby narrow the diversity of information to which users are exposed. We investigate training-data memorization as one mechanism behind this concentration. Using approximately 20 million browsing sessions from a Japanese news service, we continually trained an instruction-tuned Llama 3 8B model to generate the next article title in a session. The generated-title distribution was more concentrated and less diverse than the distribution of titles observed in the browsing logs. Frequently repeated target titles exhibited greater verbatim and approximate memorization, and more strongly memorized titles were generated in response to a wider range of prompts. These findings support a pathway in which popularity in behavioral logs becomes repeated training text, is preferentially memorized, and is subsequently reproduced by a generative recommender. An exploratory target-level deduplication condition reduced both memorization and output concentration, while substantially reducing the training-set size and recommendation accuracy. Our results suggest that popularity bias in generative recommendation may arise not only from user preferences or ranking objectives, but also from the model’s tendency to reproduce frequently encountered strings, with important implications for information diversity on news platforms.
\end{abstract}


\section{Introduction}

Large language models (LLMs) have expanded recommender systems beyond ranking a fixed candidate set. In \emph{generative recommendation}, a model directly generates an item identifier, title, or natural-language recommendation conditioned on a user's interaction history~\cite{10.1145/3678004,10.1145/3637528.3671474}. This formulation can exploit semantic knowledge learned during pretraining, accommodate textual item descriptions, support sparse-history users, and provide natural-language explanations~\cite{rajput2023recommender,10.1145/3523227.3546767,10.1145/3580488}. Recent work has further reformulated large-scale recommendation as sequential transduction and demonstrated that generative recommender architectures can scale to industrial deployments with billions of users~\cite{zhai2024actions}. The generative formulation also changes the mechanism by which exposure is allocated. A conventional ranker scores existing candidates; a generative model learns a token distribution from repeated strings and may directly emit titles that occur frequently in the training corpus.

\begin{figure*}[t]
\centering
\includegraphics[width=16cm]{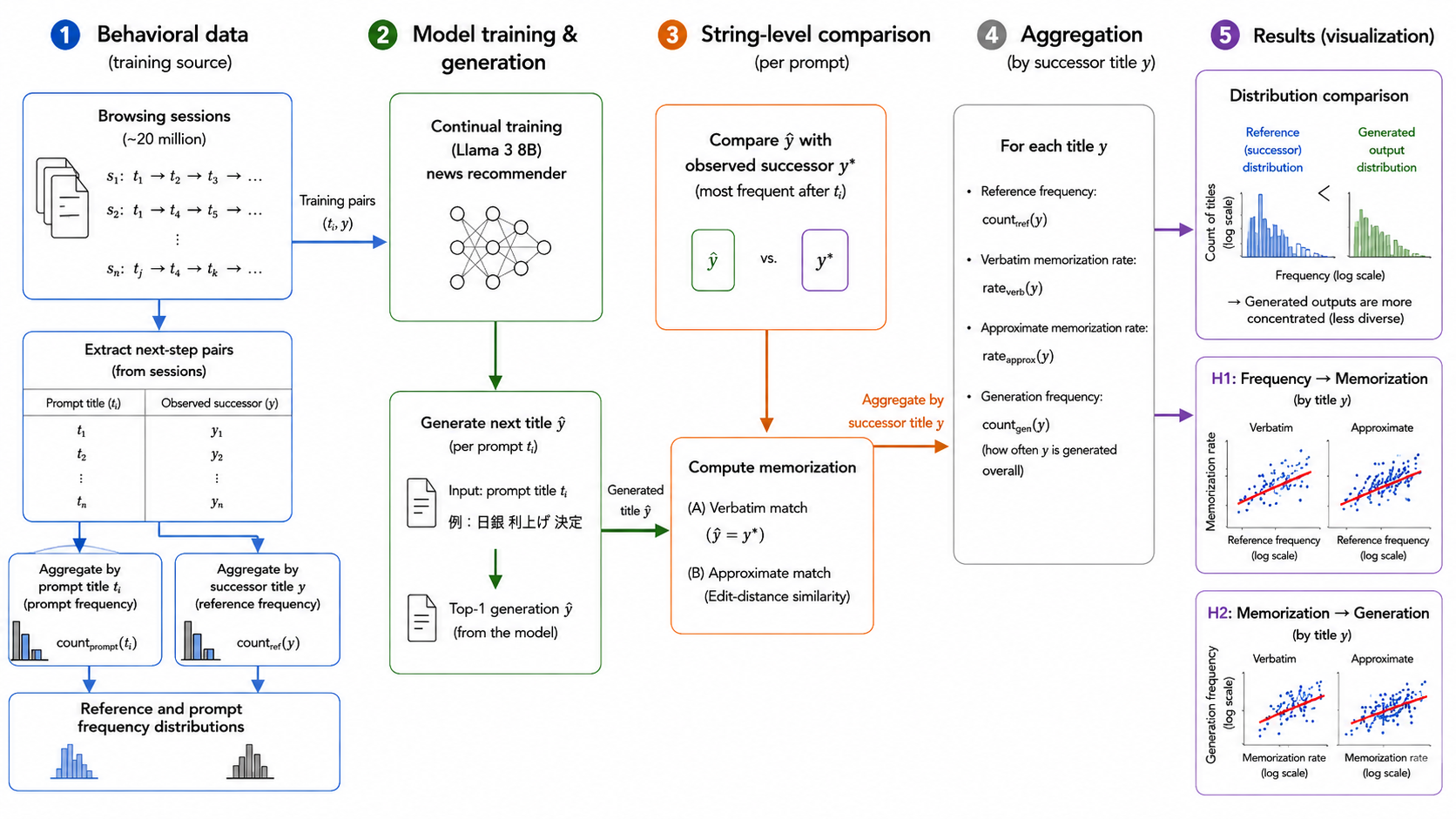}
\tightfigurecaption
\caption{Overview of the data, model, and analysis pipeline. Browsing sessions are converted into next-step prompt–successor pairs and used to continually train a Llama 3 8B generative news recommender. For each prompt, the model generates a next title, which is compared with the most frequent observed successor to measure verbatim and approximate memorization. Statistics are then aggregated by successor title to examine the concentration of generated outputs, the relationship between reference frequency and memorization (H1), and the relationship between memorization and generation frequency (H2).}
\label{fig:overview}
\end{figure*}

This distinction matters for news.
Popularity concentration in entertainment or commerce can reduce catalog coverage, but concentration in news recommendation also shapes the diversity of information encountered by citizens.
Unlike entertainment platforms (e.g., video streaming or e-commerce), where popularity concentration primarily limits catalog exploration, news recommendation carries a normative responsibility to keep citizens informed about emerging social conditions without reinforcing stereotypical views of society. When a recommender privileges already popular or historically frequent headlines, it risks recycling familiar viral stories at the expense of timeliness, freshness, and diverse civic perspectives.
Prior work has linked diversity in news consumption to engagement and has emphasized that popularity diversity is distinct from topical diversity~\cite{sonoda2022engagement}.
More broadly, recommendation feedback loops can increase homogeneity even when they optimize standard utility measures~\cite{chaney2018confounding}.
Recent evidence from agentic e-commerce likewise suggests that AI-mediated choice can concentrate demand on a small set of modal products, with outcomes varying substantially across models~\cite{allouah2026agentic}. If an LLM-based recommender simply reproduces frequently observed title transitions, the appearance of semantic intelligence may conceal a distributional mechanism that favors already dominant content.

Popularity bias is a longstanding problem in recommender systems~\cite{klimashevskaia2024popularity}. Recent studies have examined whether LLM-based recommendation exacerbates or alleviates it, with results depending on the task, model, and prompting setup~\cite{10.1145/3583780.3615077,10.1007/978-3-031-56060-6_24,lichtenberg2024popularity}. However, most analyses measure the resulting exposure pattern without identifying why a generative model concentrates its outputs. We investigate one possible mechanism: \emph{training-data memorization}, the tendency of a model to reproduce strings that appeared in training data. Memorization is known to increase with repeated examples and can be mitigated by training-data deduplication~\cite{carlini2023quantifying,pmlr-v162-kandpal22a,lee-etal-2022-deduplicating}. Recent work has begun to study memorization in recommendation data, either as leakage from general-purpose LLMs~\cite{colacicco2026memorization} or as one-hop transition reuse in generative recommenders~\cite{kim2026memorization}. We complement this work by studying string-level memorization as an explanation for output concentration in a large-scale real-world news setting.

We formulate the mechanism as one premise and two hypotheses. \textbf{Premise (P):} item occurrences in behavioral logs are highly unequal; when item titles are used as generation targets, popularity also appears as repeated strings. \textbf{Hypothesis 1 (H1):} frequently repeated target titles are more strongly memorized. \textbf{Hypothesis 2 (H2):} strongly memorized titles are emitted for more prompts, producing a concentrated output distribution. Figure~\ref{fig:overview} summarizes the analysis.

Our contributions are as follows. First, we connect popularity bias in generative recommendation to string-level memorization and operationalize the proposed pathway using frequency, verbatim memorization, and edit-distance similarity. Second, we evaluate the pathway on approximately 20 million sessions from a Japanese news service rather than on a popularity-obfuscated public benchmark. Third, we show that the generated distribution is more concentrated than the observed successor distribution and that the two links in the proposed pathway---frequency to memorization and memorization to generation frequency---are both present, although their correlations are moderate. Finally, we report an exploratory target-level deduplication condition and explicitly document its accuracy--diversity trade-off and its data-size confound.

\section{Related Work}

\subsection{Generative Recommendation and Popularity Bias}

Generative recommendation treats user--item interactions as
sequences to be modeled with language-model objectives.
Recent work distinguishes generative recommenders by how items
are tokenized, including sparse identifiers, natural-language text,
and semantic identifiers~\cite{li2025tridecoupled}.
Our setting belongs to the text-based family: article titles
themselves serve as item representations and generation targets.
This distinction is central to our analysis because item popularity
is instantiated as repeated lexical sequences, making string-level
memorization a potential mechanism of output concentration.

P5 unifies recommendation tasks as language processing~\cite{10.1145/3523227.3546767}, while generative retrieval maps user histories to item identifiers~\cite{rajput2023recommender}. HSTU reformulates large-scale recommendation as sequential transduction and reports scaling behavior and online gains in industrial deployments~\cite{zhai2024actions}. RecGPT reports that autoregressive pretraining can outperform instruction-style formulations for text-based recommendation~\cite{ngo-nguyen-2024-recgpt}, and LLM-based modeling has also been applied to article-viewing flows~\cite{10.1145/3589335.3648305}. LLMs have additionally been used as user simulators to synthesize click probabilities and train lightweight news recommenders without real interaction data~\cite{park2025laus}. Whereas that approach addresses data scarcity and privacy concerns, we study a complementary risk of adapting a generative model directly to real interaction logs: the reproduction of their frequency concentration through memorization.

Popularity bias occurs when a small set of frequently consumed items receives disproportionate exposure~\cite{klimashevskaia2024popularity}. Text matching can reduce popularity bias in sequential recommendation~\cite{10.1145/3583780.3615077}, and prompting history-aware LLM rankers can alter their bias~\cite{10.1007/978-3-031-56060-6_24}. Lichtenberg \emph{et al.} found that a simple general-purpose LLM recommender could exhibit less popularity bias than conventional baselines in a movie task~\cite{lichtenberg2024popularity}, illustrating that LLM effects are not uniform. Our setting differs because the model is continually trained on proprietary interaction logs and generates article titles without a candidate set. We ask whether this adaptation process turns frequency into memorized text and then into concentrated exposure.

\subsection{Memorization in Language Models and Recommenders}

Language-model memorization is commonly measured by whether training strings can be reproduced exactly or approximately~\cite{ishihara-2023-training}. Across model families, repeated sequences are more likely to be memorized~\cite{carlini2023quantifying}; deduplication reduces both memorization and privacy risks~\cite{pmlr-v162-kandpal22a,lee-etal-2022-deduplicating}. Similar tendencies have been observed in Japanese language models and Japanese news corpora~\cite{ishihara-takahashi-2024-quantifying-memorization,kiyomaru-etal-2024-comprehensive-analysis}.

Recommendation-specific memorization has recently attracted attention. Colacicco \emph{et al.} test whether general-purpose LLMs reveal MovieLens information~\cite{colacicco2026memorization}. Kim \emph{et al.} analyze one-hop memorization in LLM-based generative recommendation and show that much of the performance gain occurs where a target is a direct successor observed during training~\cite{kim2026memorization}. Their focus is generalization and recommendation accuracy. Our focus is the social distribution of generated exposure: whether string memorization helps explain why many distinct prompts collapse onto the same popular outputs.

\subsection{News Diversity as a Social Outcome}

News recommendation is not evaluated solely by accuracy. Sonoda \emph{et al.} distinguish content diversity from popularity diversity and show that both are related to engagement in a news application~\cite{sonoda2022engagement}. Work on algorithmic confounding further demonstrates how repeated retraining on behavior shaped by earlier recommendations can increase homogeneity~\cite{chaney2018confounding}. Related evidence from AI shopping agents shows that model-mediated decisions may concentrate demand on a few modal products, although that work attributes concentration to agent decision-making, interface position, and model-specific preferences rather than memorization~\cite{allouah2026agentic}. These findings motivate our emphasis on aggregate output concentration. We do not claim that a single next-title model directly causes political polarization or filter bubbles; rather, we identify a technical pathway that could narrow the set of articles exposed at scale if deployed without diversity controls.

\section{Data and Generative News Recommender}
\label{sec:model}

\subsection{Browsing-Session Data}
\label{sec:data}

We use browsing logs from Nikkei Online, a Japanese subscription news service. Public recommendation datasets may transform or anonymize business metrics in ways that obscure true item frequencies~\cite{10.1145/3383313.3412256}. Because the proposed mechanism depends on frequency, we analyze the original within-service distribution. The corpus contains approximately 20 million sessions with at least three viewed articles during a fixed observation window in January 2023.

For a session containing titles $a_1,\ldots,a_N$, we construct the text
\begin{equation}
 a_1\ \articlesep\ a_2\ \articlesep\ \cdots\ a_N.
\end{equation}
The special token \articlesep{} denotes a transition between articles. Multiple session strings are concatenated with a separator token, subject to a maximum length of 512 tokens. This autoregressive formulation follows evidence that language-model pretraining can be effective for text-based recommendation~\cite{ngo-nguyen-2024-recgpt}.

The logs and trained model are proprietary and cannot be released. The paper reports aggregate distributions and selected
article-title examples; it does not report user identifiers
or individual browsing histories.

\subsection{Continual Training}

We continually train the instruction-tuned \texttt{Meta-Llama-3-8B-Instruct} model on the session corpus using Low-Rank Adaptation (LoRA)~\cite{hu2022lora}. We refer to the resulting model as \model. The LoRA rank is 16, training lasts two epochs, and checkpoints are stored every 1,000 steps. The learning rate is $10^{-4}$ and the maximum sequence length is 512. Training uses Python 3.10, Transformers 4.36.0, and PyTorch 2.1.0. Computation is performed on eight \texttt{ml.p4d.24xlarge} instances, each equipped with eight NVIDIA A100 GPUs.

\subsection{Generation and Evaluation Pairs}

Given a prompt of the form ``article title \articlesep,'' \model{} generates the next title. We use greedy decoding because the goal is to characterize the model's highest-probability continuation rather than stochastic recommendation diversity. Generation stops after at most 50 tokens.

For the analysis, we restrict attention to the first transition in each session. The first viewed title is the prompt and the second is the successor. Because the same first title may be followed by different articles, we retain its most frequent successor as the reference. This yields 193,815 distinct prompt titles and 101,026 distinct reference titles. We generate one continuation for each prompt and extract the first predicted article title. The resulting outputs contain 20,689 distinct title strings.

\section{Measures}
\label{sec:measures}

\subsection{Popularity Concentration and Diversity}

We characterize the prompt, reference, and generated-title distributions with four aggregate measures.

\textbf{Frequency concentration.} We report a Gini-based concentration measure derived from the item-frequency distribution. Larger values indicate that occurrences are concentrated on fewer titles. We also visualize Lorenz curves.

\textbf{Entropy.} We report the average information content of the empirical item distribution, i.e., its Shannon entropy. Larger values indicate a more even frequency distribution.

\textbf{GS score.} Following Waller and Anderson~\cite{waller2019generalists}, we embed article titles and compute a semantic concentration score. We use the Japanese static embedding model \texttt{hotchpotch/static-embedding-japanese}. The score ranges from $-1$ to 1; larger values indicate that the titles are semantically more similar and thus less diverse.

\textbf{Catalog coverage.} We count the number of distinct title strings in each distribution.

\subsection{String-Level Memorization}

For generated title $g$ and reference title $y$, we use two measures adapted from prior Japanese memorization analysis~\cite{ishihara-takahashi-2024-quantifying-memorization}.

\textbf{Verbatim memorization} is the number of leading characters shared by $g$ and $y$:
\begin{equation}
M_{\mathrm{verb}}(g,y)=\max\{k: g_{1:k}=y_{1:k}\}.
\end{equation}

\textbf{Approximate memorization} is normalized Levenshtein similarity:
\begin{equation}
M_{\mathrm{approx}}(g,y)=1-\frac{d_{\mathrm{lev}}(g,y)}{\max(|g|,|y|)}.
\end{equation}
Both measures increase as the generated continuation more closely reproduces the observed successor.

We define \emph{reference frequency} as the number of times a title
appears as the second article in the approximately 20 million
original sessions, and \emph{generation frequency} as its count in the generated-output set. Because both distributions are heavy-tailed, frequency is log-transformed for correlation analysis. Pearson's $r$ is used to quantify the two hypothesized links. With this large sample, statistical significance is not informative by itself; we therefore emphasize effect sizes and distributional patterns.

\section{Results}
\label{sec:results}

\subsection{P: Generated Recommendations Are More Concentrated}

\begin{figure}[t]
\centering
\includegraphics[width=0.96\columnwidth]{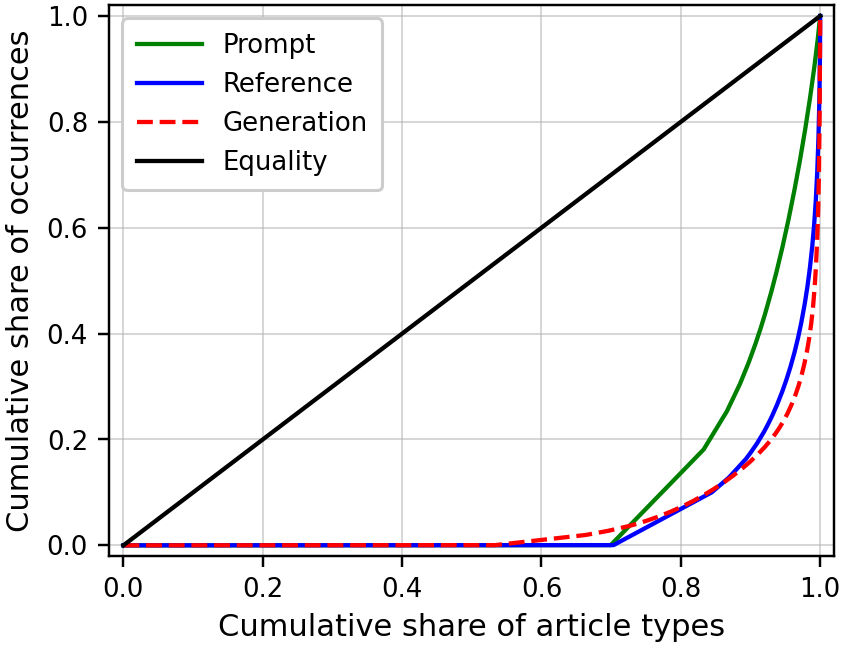}
\tightfigurecaption
\caption{Lorenz curves for prompt, reference, and generated-title occurrence distributions. The generated outputs are more concentrated than the observed references.}
\label{fig:lorenz}
\end{figure}

Figure~\ref{fig:lorenz} shows Lorenz curves for prompts, references, and generated titles. All three distributions are unequal, but the generated distribution is the most concentrated. Table~\ref{tab:distribution} quantifies the difference. Relative to the references, the generated outputs have a larger concentration value (0.9072 versus 0.9003), lower entropy (7.169 versus 9.147), a larger GS score (0.3364 versus 0.1911), and approximately one fifth as many distinct titles (20,689 versus 101,026). Thus, the model maps many prompts to a substantially narrower set of titles than users actually viewed as successors.

The Pearson correlation between prompt frequency and generation frequency is 0.2905. Popular prompts therefore tend to produce popular outputs to some extent, but prompt popularity alone does not explain the output concentration. We next examine target repetition and memorization.

\begin{table}[t]
\centering
\caption{Distributional measures. Larger concentration and GS values indicate less diversity; larger entropy and type count indicate more diversity.}
\label{tab:distribution}
\setlength{\tabcolsep}{4pt}
\begin{tabular}{lrrr}
\toprule
Measure & Prompt & Reference & Generation \\
\midrule
Concentration & 0.8180 & 0.9003 & 0.9072 \\
Entropy & 10.710 & 9.147 & 7.169 \\
GS score & 0.1736 & 0.1911 & 0.3364 \\
Distinct titles & 193,815 & 101,026 & 20,689 \\
\bottomrule
\end{tabular}
\end{table}

\subsection{H1: Repeated References Are More Memorized}

\begin{figure*}[t]
\centering
\begin{minipage}{0.48\textwidth}
\centering
\includegraphics[width=\linewidth]{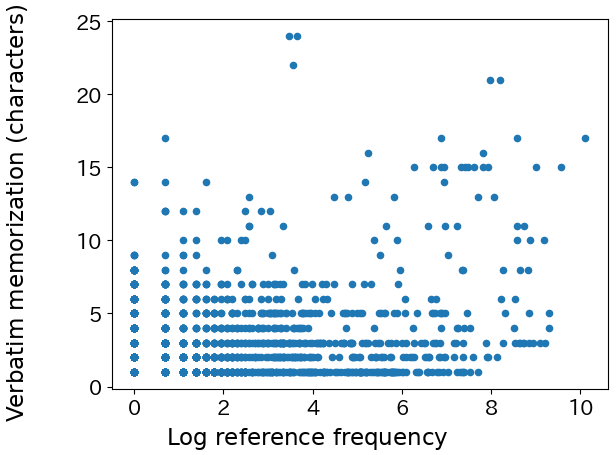}
\end{minipage}
\hfill
\begin{minipage}{0.48\textwidth}
\centering
\includegraphics[width=\linewidth]{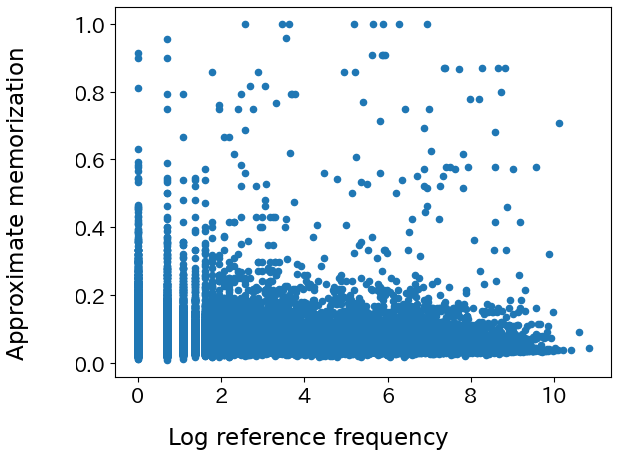}
\end{minipage}
\tightfigurecaption
\caption{Relationship between reference frequency and memorization. Reference frequency is the number of times a title appears as the second article in the approximately 20 million original sessions. Its logarithm correlates with verbatim memorization, measured as the number of matching initial characters ($r=0.3042$; left), and with approximate memorization, measured by normalized Levenshtein similarity ($r=0.3344$; right). Cases with zero verbatim memorization are omitted from the left panel.}
\label{fig:h1}
\end{figure*}

Figure~\ref{fig:h1} relates reference frequency to the two memorization measures. Log reference frequency correlates with verbatim memorization at $r=0.3042$ and with approximate memorization at $r=0.3344$. The relationships are moderate and noisy, but they are consistent for both verbatim and approximate memorization: titles that appear more often as second articles in the original sessions are more likely to be reproduced by the trained model.

We also examine prompt length because longer prompts have been associated with greater discoverable memorization in general language models~\cite{carlini2023quantifying}. In this setting, title length has essentially no positive relationship with verbatim memorization ($r=0.0167$) and a weak negative relationship with approximate memorization ($r=-0.1195$). The frequency effect is therefore not reducible to longer title prompts.

\subsection{H2: Memorized Titles Are Generated More Often}

\begin{figure*}[t]
\centering
\begin{minipage}{0.48\textwidth}
\centering
\includegraphics[width=\linewidth]{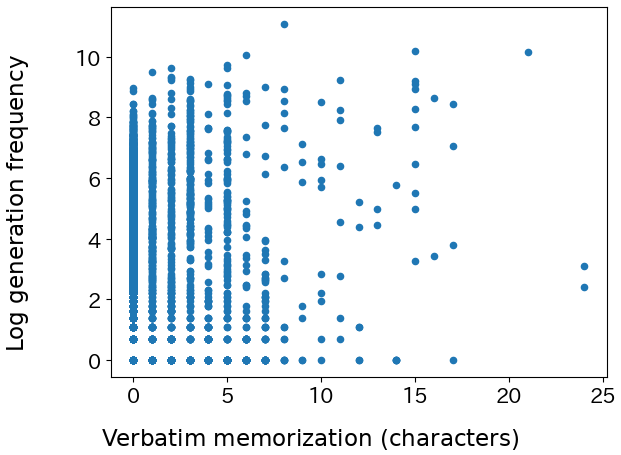}
\end{minipage}
\hfill
\begin{minipage}{0.48\textwidth}
\centering
\includegraphics[width=\linewidth]{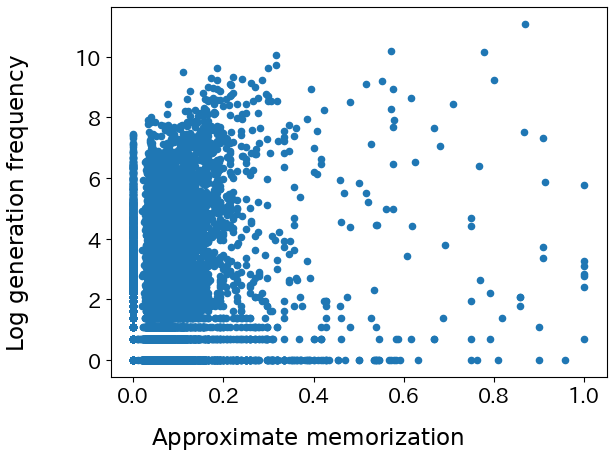}
\end{minipage}
\tightfigurecaption
\caption{Relationship between memorization and output frequency. The logarithm of generation frequency correlates with verbatim memorization ($r=0.1965$; left) and approximate memorization ($r=0.3084$; right). Cases with zero verbatim memorization are omitted from the left panel.}
\label{fig:h2}
\end{figure*}

Figure~\ref{fig:h2} relates memorization to generation frequency. Verbatim memorization correlates with log generation frequency at $r=0.1965$; approximate memorization correlates at $r=0.3084$. The latter is similar in magnitude to the frequency--memorization relationship in H1. The plots also show that titles with high verbatim memorization are rarely confined to a single prompt: many become repeated outputs.

Table~\ref{tab:examples} gives five examples with high verbatim memorization.
We retain the original Japanese titles because memorization is
measured at the character level. The first three concern a three-part golf series. The model reproduces the title template and sometimes maps multiple parts to the same continuation. In the last two rows, two unrelated prompts share the same popular successor in the reference set and produce the same partially corrupted continuation. That generated title occurs 25,840 times. These cases illustrate how a title learned from many transitions can become a default continuation for a wide range of prompts.

\begin{table*}[t]
\centering
\caption{Examples with the highest verbatim memorization.
The highlighted spans indicate verbatim matches between the
reference and generated titles.
The original Japanese titles are retained because memorization
is measured at the character level.}
\label{tab:examples}

\footnotesize
\setlength{\tabcolsep}{2pt}
\setlength{\fboxsep}{1pt}
\renewcommand{\arraystretch}{1.15}

\begin{tabular}{
    p{0.19\textwidth}
    r
    p{0.19\textwidth}
    p{0.21\textwidth}
    r
    r
    r
}
\hline
Prompt
& Ref. freq.
& Reference
& Generated title
& Gen. freq.
& Verbatim
& Approx. \\
\hline

90切りの現実考察とそれを可能にする練習法（中）
& 10
& 90切りの現実考察とそれを可能にする練習法（上）
& \colorbox{green!20}{\strut 90切りの現実考察とそれを可}
  \colorbox{green!20}{\strut 能にする練習法（上）}
& 22
& 24
& 1.000 \\ \hline

90切りの現実考察とそれを可能にする練習法（上）
& 11
& 90切りの現実考察とそれを可能にする練習法（下）
& \colorbox{green!20}{\strut 90切りの現実考察とそれを可}
  \colorbox{green!20}{\strut 能にする練習法（下）}
& 11
& 24
& 1.000 \\ \hline

90切りの現実考察とそれを可能にする練習法（下）
& 12
& 90切りの現実考察とそれを可能にする練習法（中）
& \colorbox{green!20}{\strut 90切りの現実考察とそれを可}
  \colorbox{green!20}{\strut 能にする練習法}（上）
& 22
& 22
& 0.958 \\ \hline

中国、制限なし春節スタート 旅客倍増で感染拡大警戒
& 141
& その決断、本当に自分の意思? 選択より前に脳が反応
& \colorbox{green!20}{\strut その決断、本当に自分の意思?}
  \colorbox{green!20}{\strut 選択より前に}「アジッド」
& 25840
& 21
& 0.777 \\ \hline

Twitter社員、買収前の5分の1に 1300人に減少と米報道
& 3103
& その決断、本当に自分の意思? 選択より前に脳が反応
& \colorbox{green!20}{\strut その決断、本当に自分の意思?}
  \colorbox{green!20}{\strut 選択より前に}「アジッド」
& 25840
& 21
& 0.777 \\ \hline

\end{tabular}
\end{table*}

\subsection{Interpretation}

The evidence is correlational rather than a complete causal decomposition, but the three distributions and two links form a coherent account. First, observed next-title frequencies are highly unequal. Second, frequently repeated reference strings are more closely reproduced. Third, titles with stronger reproduction are emitted for more prompts. The model therefore appears to convert behavioral frequency into a lexical prior that can dominate generation.

This mechanism differs subtly from simply recommending popular items because they are globally popular. The model is trained on local title transitions, and the same successor can be associated with many different prompts. Memorization allows that successor string to become a high-probability continuation across contexts. The result is a many-to-one collapse that reduces catalog coverage and semantic diversity.

\section{Exploratory Target-Level Deduplication}
\label{sec:dedup}

The proposed mechanism suggests that reducing repeated targets may reduce both memorization and output concentration. We conduct an exploratory intervention by restricting each session to its first two titles and selecting sessions such that no target title is duplicated. We train \dedupmodel{} with the same stated hyperparameters as the original model. The resulting training set contains 193,860 sessions, approximately one hundredth of the original corpus.

Table~\ref{tab:dedup} reports the comparison. Verbatim memorization falls from 0.1356 to 0.0164 on average, approximate memorization falls from 0.0360 to 0.0264, and the correlations between memorization and generation frequency become much smaller. The number of generated title types nearly doubles from 20,689 to 39,479. Frequency-based diversity also improves: concentration declines from 0.9072 to 0.8901 and entropy rises from 7.169 to 7.222. However, the GS score increases from 0.3364 to 0.4658, indicating that semantic diversity does not improve under this measure. Exact-match accuracy also drops sharply, from 2,461 to 81 correct predictions.

\begin{table*}[t]
\centering
\scriptsize
\caption{Original and target-deduplicated training conditions. The intervention changes both duplication and data volume, so differences cannot be attributed to deduplication alone.}
\label{tab:dedup}
\setlength{\tabcolsep}{4pt}
\begin{tabular}{lrrrrrrrrr}
\toprule
Condition & Conc. & Entropy & GS & Types & \shortstack{Mean verbatim\\memorization} & \shortstack{Mean approximate\\memorization} & \shortstack{$r$: verbatim\\memorization} & \shortstack{$r$: approximate\\memorization} & Exact-match correct \\
\midrule
Original & 0.9072 & 7.169 & 0.3364 & 20,689 & 0.1356 & 0.0360 & 0.1965 & 0.3084 & 2,461 \\
Target-deduplicated & 0.8901 & 7.222 & 0.4658 & 39,479 & 0.0164 & 0.0264 & 0.0692 & 0.0395 & 81 \\
\bottomrule
\end{tabular}
\end{table*}

The intervention should not be interpreted as isolating a causal effect of deduplication. It simultaneously flattens the target-frequency distribution, removes many training examples, and greatly reduces the total number of optimization updates per epoch. A matched-size random subsample with duplicates retained would be required to separate these explanations. Because the original logs needed for that control are no longer available for this study, we treat the result as a feasibility case study: frequency-flattened training can reduce memorization and broaden output coverage, but may also sacrifice predictive accuracy and does not necessarily improve semantic diversity.

\balance
\section{Discussion}

\subsection{Implications for Information Ecosystems}

The results suggest that an LLM-based news recommender may reproduce popularity through a mechanism that is partly hidden by natural-language generation. A generated title can appear contextually plausible while functioning as a memorized default. If such outputs are repeatedly exposed and subsequently enter future training logs, the process could form a feedback loop in which historical popularity is converted into model probability and then into new exposure. Our experiment does not observe such a deployed loop, but prior work shows that recommendation feedback can increase homogeneity~\cite{chaney2018confounding}. This makes memorization relevant not only to privacy and copyright, but also to the distribution of attention.

For news providers, output coverage and popularity diversity should therefore be evaluated alongside next-item accuracy.
The exact-match decline in the deduplicated condition illustrates the accuracy–diversity trade-off~\cite{zhang2023cdn}.
To align generative news recommendation with journalistic norms—delivering timely updates on emerging events without reproducing stereotypical or stale head articles—production systems should combine three methodological adjustments rather than relying on indiscriminate deduplication: (1) frequency-aware loss reweighting with recency decay so older viral headlines do not dominate lexical priors; (2) candidate-grounded or constrained decoding (e.g., trie-based or retrieval-augmented constraints) to guarantee that generated strings map strictly to valid, current catalog articles without suffix corruption; and (3) diversity-aware reranking enforcing editorial exposure floors across topics and viewpoints. Beyond this single snapshot, our insider auditing framework—linking training-log duplication directly to decoding behavior—can be extended to track temporal freshness decay, compare sparse/semantic-ID tokenization against title generation, and monitor deployed feedback loops in real time.

\subsection{Limitations and Required Validation}
\label{sec:limitations}

First, the study uses one Japanese news service, one observation period, one 8B model, and greedy decoding. Moreover, our setting is a text-based generative recommender in which article titles themselves serve as natural-language item representations and generation targets. The proposed mechanism may therefore depend on the item-tokenization scheme: our character-level memorization measures do not directly apply to recommenders based on sparse or semantic identifiers, and whether analogous popularity--memorization relationships arise under those representations remains an open question. The absolute concentration values may also differ across domains, model scales, languages, and stochastic decoding strategies.

Second, the analysis is correlational. Although frequency, memorization, and output concentration align with the hypothesized pathway, unobserved properties of article titles or transitions may affect all three. Third, the evaluation uses a modal successor for each prompt and therefore does not represent the full conditional distribution of user behavior. Fourth, the target-deduplication experiment has a major data-size confound, as discussed above. Fifth, the raw logs and model cannot be released, limiting independent reproduction. We therefore report aggregate statistics and detailed metric definitions, and future work should release evaluation code or equivalent reproducibility materials where possible. Sixth, the paper studies aggregate item exposure rather than fairness across user groups or viewpoints. It should not be read as evidence of demographic unfairness or ideological polarization.

\subsection{Ethics and Data Access}
The study uses proprietary behavioral logs under Nikkei's internal data-governance procedures for research use. User and session identifiers were removed or pseudonymized before analysis. The study analyzes observational service logs and involved no intervention with users. Only aggregate statistics and selected article-title examples are reported; no user identifiers or individual session traces are included in the manuscript. Raw logs and trained weights are not distributed.

\section{Conclusion}

We analyzed popularity concentration in an LLM-based generative news recommender through the lens of training-data memorization. Generated titles were markedly less diverse than observed successors. Repeated reference titles were more strongly memorized, and memorized titles were generated for more prompts, supporting a mechanism in which behavioral frequency is transformed into repeated lexical output. An exploratory target-level deduplication condition reduced frequency concentration and memorization but also reduced the training set by two orders of magnitude, worsened exact-match accuracy, and did not improve semantic diversity. The central contribution is therefore not a definitive mitigation method, but an empirical account of how generative recommendation can reproduce popularity. Evaluating this mechanism is important when such systems mediate access to socially consequential information.

\section*{Acknowledgment}
The author thanks Kenta Yamada for assistance with data collection, Kazumasa Omura for detailed comments on the manuscript, and Keisuke Yanagi for advice from the recommender systems perspective.

\bibliographystyle{IEEEtran}
\bibliography{refs}

@inproceedings{ngo-nguyen-2024-recgpt,
    author = "Ngo, Hoang and Nguyen, Dat Quoc",
    title = "{R}ec{GPT}: Generative Pre-training for Text-based Recommendation",
    booktitle = "Proceedings of the 62nd Annual Meeting of the Association for Computational Linguistics (Volume 2: Short Papers)",
    month = aug,
    year = "2024",
    address = "Bangkok, Thailand",
    publisher = "Association for Computational Linguistics",
    pages = "302--313",
    doi = "10.18653/v1/2024.acl-short.29",
    url = "https://aclanthology.org/2024.acl-short.29/",
}

@inproceedings{10.1145/3589335.3648305,
    author = {Liu, Zhenghao and Chen, Zulong and Zhang, Moufeng and Duan, Shaoyang and Wen, Hong and Li, Liangyue and Li, Nan and Gu, Yu and Yu, Ge},
    title = {Modeling User Viewing Flow using Large Language Models for Article Recommendation},
    booktitle = {Companion Proceedings of the ACM Web Conference 2024},
    series = {WWW '24},
    year = {2024},
    address = {New York, NY, USA},
    publisher = {Association for Computing Machinery},
    pages = {83--92},
    doi = {10.1145/3589335.3648305},
    url = {https://doi.org/10.1145/3589335.3648305},
}

@inproceedings{10.1145/3637528.3671474,
    author = {Deldjoo, Yashar and He, Zhankui and McAuley, Julian and Korikov, Anton and Sanner, Scott and Ramisa, Arnau and Vidal, Ren{\'e} and Sathiamoorthy, Maheswaran and Kasirzadeh, Atoosa and Milano, Silvia},
    title = {A Review of Modern Recommender Systems Using Generative Models (Gen-RecSys)},
    booktitle = {Proceedings of the 30th ACM SIGKDD Conference on Knowledge Discovery and Data Mining},
    series = {KDD '24},
    year = {2024},
    address = {New York, NY, USA},
    publisher = {Association for Computing Machinery},
    pages = {6448--6458},
    doi = {10.1145/3637528.3671474},
    url = {https://doi.org/10.1145/3637528.3671474},
}

@article{10.1145/3678004,
    author = {Lin, Jianghao and Dai, Xinyi and Xi, Yunjia and Liu, Weiwen and Chen, Bo and Zhang, Hao and Liu, Yong and Wu, Chuhan and Li, Xiangyang and Zhu, Chenxu and Guo, Huifeng and Yu, Yong and Tang, Ruiming and Zhang, Weinan},
    title = {How Can Recommender Systems Benefit from Large Language Models: A Survey},
    journal = {ACM Trans. Inf. Syst.},
    volume = {43},
    number = {2},
    pages = {1--47},
    month = mar,
    year = {2025},
    publisher = {Association for Computing Machinery},
    address = {New York, NY, USA},
    issn = {1046-8188},
    doi = {10.1145/3678004},
    url = {https://doi.org/10.1145/3678004},
}

@inproceedings{10.1007/978-3-031-56060-6_24,
    author = {Hou, Yupeng and Zhang, Junjie and Lin, Zihan and Lu, Hongyu and Xie, Ruobing and McAuley, Julian and Zhao, Wayne Xin},
    title = {Large Language Models are Zero-Shot Rankers for Recommender Systems},
    booktitle = {Advances in Information Retrieval: 46th European Conference on Information Retrieval, ECIR 2024, Glasgow, UK, March 24--28, 2024, Proceedings, Part II},
    year = {2024},
    address = {Berlin, Heidelberg},
    publisher = {Springer-Verlag},
    pages = {364--381},
    doi = {10.1007/978-3-031-56060-6_24},
    url = {https://doi.org/10.1007/978-3-031-56060-6_24},
}

@inproceedings{10.1145/3583780.3615077,
    author = {Liu, Zhenghao and Mei, Sen and Xiong, Chenyan and Li, Xiaohua and Yu, Shi and Liu, Zhiyuan and Gu, Yu and Yu, Ge},
    title = {Text Matching Improves Sequential Recommendation by Reducing Popularity Biases},
    booktitle = {Proceedings of the 32nd ACM International Conference on Information and Knowledge Management},
    series = {CIKM '23},
    year = {2023},
    address = {New York, NY, USA},
    publisher = {Association for Computing Machinery},
    pages = {1534--1544},
    doi = {10.1145/3583780.3615077},
    url = {https://doi.org/10.1145/3583780.3615077},
}

@inproceedings{hu2022lora,
    author = {Hu, Edward J. and Shen, Yelong and Wallis, Phillip and Allen-Zhu, Zeyuan and Li, Yuanzhi and Wang, Shean and Wang, Lu and Chen, Weizhu},
    title = {{LoRA}: Low-Rank Adaptation of Large Language Models},
    booktitle = {International Conference on Learning Representations},
    year = {2022},
    url = {https://openreview.net/forum?id=nZeVKeeFYf9},
}

@inproceedings{rajput2023recommender,
    author = {Rajput, Shashank and Mehta, Nikhil and Singh, Anima and Keshavan, Raghunandan Hulikal and Vu, Trung and Heldt, Lukasz and Hong, Lichan and Tay, Yi and Tran, Vinh Q. and Samost, Jonah and Kula, Maciej and Chi, Ed H. and Sathiamoorthy, Maheswaran},
    title = {Recommender Systems with Generative Retrieval},
    booktitle = {Advances in Neural Information Processing Systems},
    volume = {36},
    year = {2023},
    url = {https://openreview.net/forum?id=BJ0fQUU32w},
}

@inproceedings{10.1145/3523227.3546767,
    author = {Geng, Shijie and Liu, Shuchang and Fu, Zuohui and Ge, Yingqiang and Zhang, Yongfeng},
    title = {Recommendation as Language Processing ({RLP}): A Unified Pretrain, Personalized Prompt \& Predict Paradigm ({P5})},
    booktitle = {Proceedings of the 16th ACM Conference on Recommender Systems},
    series = {RecSys '22},
    year = {2022},
    address = {New York, NY, USA},
    publisher = {Association for Computing Machinery},
    pages = {299--315},
    doi = {10.1145/3523227.3546767},
    url = {https://doi.org/10.1145/3523227.3546767},
}

@article{10.1145/3580488,
    author = {Li, Lei and Zhang, Yongfeng and Chen, Li},
    title = {Personalized Prompt Learning for Explainable Recommendation},
    journal = {ACM Trans. Inf. Syst.},
    volume = {41},
    number = {4},
    articleno = {103},
    numpages = {26},
    month = mar,
    year = {2023},
    address = {New York, NY, USA},
    publisher = {Association for Computing Machinery},
    issn = {1046-8188},
    doi = {10.1145/3580488},
    url = {https://doi.org/10.1145/3580488},
}

@inproceedings{ishihara-2023-training,
    author = "Ishihara, Shotaro",
    title = "Training Data Extraction From Pre-trained Language Models: A Survey",
    booktitle = "Proceedings of the 3rd Workshop on Trustworthy Natural Language Processing (TrustNLP 2023)",
    month = jul,
    year = "2023",
    address = "Toronto, Canada",
    publisher = "Association for Computational Linguistics",
    pages = "260--275",
    doi = "10.18653/v1/2023.trustnlp-1.23",
    url = "https://aclanthology.org/2023.trustnlp-1.23/",
}

@inproceedings{carlini2023quantifying,
    author = {Carlini, Nicholas and Ippolito, Daphne and Jagielski, Matthew and Lee, Katherine and Tramer, Florian and Zhang, Chiyuan},
    title = {Quantifying Memorization Across Neural Language Models},
    booktitle = {The Eleventh International Conference on Learning Representations},
    year = {2023},
    url = {https://openreview.net/forum?id=TatRHT_1cK},
}

@inproceedings{ishihara-takahashi-2024-quantifying-memorization,
    author = "Ishihara, Shotaro and Takahashi, Hiromu",
    title = "Quantifying Memorization and Detecting Training Data of Pre-trained Language Models using {J}apanese Newspaper",
    booktitle = "Proceedings of the 17th International Natural Language Generation Conference",
    month = sep,
    year = "2024",
    address = "Tokyo, Japan",
    publisher = "Association for Computational Linguistics",
    pages = "165--179",
    url = "https://aclanthology.org/2024.inlg-main.14",
}

@inproceedings{kiyomaru-etal-2024-comprehensive-analysis,
    author = "Kiyomaru, Hirokazu and Sugiura, Issa and Kawahara, Daisuke and Kurohashi, Sadao",
    title = "A Comprehensive Analysis of Memorization in Large Language Models",
    booktitle = "Proceedings of the 17th International Natural Language Generation Conference",
    month = sep,
    year = "2024",
    address = "Tokyo, Japan",
    publisher = "Association for Computational Linguistics",
    pages = "584--596",
    url = "https://aclanthology.org/2024.inlg-main.45",
}

@inproceedings{lee-etal-2022-deduplicating,
    author = "Lee, Katherine and Ippolito, Daphne and Nystrom, Andrew and Zhang, Chiyuan and Eck, Douglas and Callison-Burch, Chris and Carlini, Nicholas",
    title = "Deduplicating Training Data Makes Language Models Better",
    booktitle = "Proceedings of the 60th Annual Meeting of the Association for Computational Linguistics (Volume 1: Long Papers)",
    month = may,
    year = "2022",
    address = "Dublin, Ireland",
    publisher = "Association for Computational Linguistics",
    pages = "8424--8445",
    doi = "10.18653/v1/2022.acl-long.577",
    url = "https://aclanthology.org/2022.acl-long.577/",
}

@inproceedings{pmlr-v162-kandpal22a,
    author = {Kandpal, Nikhil and Wallace, Eric and Raffel, Colin},
    title = {Deduplicating Training Data Mitigates Privacy Risks in Language Models},
    booktitle = {Proceedings of the 39th International Conference on Machine Learning},
    series = {Proceedings of Machine Learning Research},
    volume = {162},
    editor = {Chaudhuri, Kamalika and Jegelka, Stefanie and Song, Le and Szepesvari, Csaba and Niu, Gang and Sabato, Sivan},
    month = jul,
    year = {2022},
    publisher = {PMLR},
    pages = {10697--10707},
    url = {https://proceedings.mlr.press/v162/kandpal22a.html},
}

@inproceedings{10.1145/3383313.3412256,
    author = {Seki, Yoshifumi and Maehara, Takanori},
    title = {A Method to Anonymize Business Metrics to Publishing Implicit Feedback Datasets},
    booktitle = {Proceedings of the 14th ACM Conference on Recommender Systems},
    series = {RecSys '20},
    year = {2020},
    address = {New York, NY, USA},
    publisher = {Association for Computing Machinery},
    pages = {4--12},
    doi = {10.1145/3383313.3412256},
    url = {https://doi.org/10.1145/3383313.3412256},
}

@article{klimashevskaia2024popularity,
    author = {Klimashevskaia, Anastasiia and Jannach, Dietmar and Elahi, Mehdi and Trattner, Christoph},
    title = {A Survey on Popularity Bias in Recommender Systems},
    journal = {User Modeling and User-Adapted Interaction},
    volume = {34},
    number = {5},
    year = {2024},
    doi = {10.1007/s11257-024-09406-0},
    url = {https://doi.org/10.1007/s11257-024-09406-0},
}

@inproceedings{waller2019generalists,
    author = {Waller, Isaac and Anderson, Ashton},
    title = {Generalists and Specialists: Using Community Embeddings to Quantify Activity Diversity in Online Platforms},
    booktitle = {Proceedings of The Web Conference 2019},
    series = {WWW '19},
    year = {2019},
    address = {New York, NY, USA},
    publisher = {Association for Computing Machinery},
    doi = {10.1145/3308558.3313729},
    url = {https://doi.org/10.1145/3308558.3313729},
}

@inproceedings{zhang2023cdn,
    author = {Zhang, Yin and Wang, Ruoxi and Cheng, Derek Zhiyuan and Yao, Tiansheng and Yi, Xinyang and Hong, Lichan and Caverlee, James and Chi, Ed H.},
    title = {Empowering Long-Tail Item Recommendation through Cross Decoupling Network},
    booktitle = {Proceedings of the 29th ACM SIGKDD Conference on Knowledge Discovery and Data Mining},
    series = {KDD '23},
    year = {2023},
    address = {New York, NY, USA},
    publisher = {Association for Computing Machinery},
    doi = {10.1145/3580305.3599814},
    url = {https://doi.org/10.1145/3580305.3599814},
}

@article{sonoda2022engagement,
    author = {Sonoda, Atom and Seki, Yoshifumi and Toriumi, Fujio},
    title = {Analyzing User Engagement in News Application Considering Popularity Diversity and Content Diversity},
    journal = {Journal of Computational Social Science},
    volume = {5},
    number = {2},
    pages = {1595--1614},
    year = {2022},
    doi = {10.1007/s42001-022-00179-3},
    url = {https://doi.org/10.1007/s42001-022-00179-3},
}

@inproceedings{chaney2018confounding,
    author = {Chaney, Allison J. B. and Stewart, Brandon M. and Engelhardt, Barbara E.},
    title = {How Algorithmic Confounding in Recommendation Systems Increases Homogeneity and Decreases Utility},
    booktitle = {Proceedings of the 12th ACM Conference on Recommender Systems},
    series = {RecSys '18},
    year = {2018},
    address = {New York, NY, USA},
    publisher = {Association for Computing Machinery},
    pages = {224--232},
    doi = {10.1145/3240323.3240370},
    url = {https://doi.org/10.1145/3240323.3240370},
}

@article{lichtenberg2024popularity,
    author = {Lichtenberg, Jan Malte and Buchholz, Alexander and Schw{\"o}bel, Pola},
    title = {Large Language Models as Recommender Systems: A Study of Popularity Bias},
    journal = {arXiv preprint arXiv:2406.01285},
    year = {2024},
    url = {https://arxiv.org/abs/2406.01285},
}

@article{kim2026memorization,
    author = {Kim, Sunwoo and Lee, Sunkyung and Ju, Clark Mingxuan and Loveland, Donald and Kumar, Bhuvesh and Shin, Kijung and Shah, Neil and Collins, Liam},
    title = {On the Memorization Behavior of Large Language Models in Generative Recommendation: Observations, Implications, and Training Strategies},
    journal = {arXiv preprint arXiv:2606.17276},
    year = {2026},
    url = {https://arxiv.org/abs/2606.17276},
}

@article{colacicco2026memorization,
    author = {Colacicco, Antonio and Guida, Vito and Di Palma, Dario and Narducci, Fedelucio and Di Noia, Tommaso},
    title = {Exploring Approaches for Detecting Memorization of Recommender System Data in Large Language Models},
    journal = {arXiv preprint arXiv:2601.02002},
    year = {2026},
    url = {https://arxiv.org/abs/2601.02002},
}

@inproceedings{zhai2024actions,
    author = {Zhai, Jiaqi and Liao, Lucy and Liu, Xing and Wang, Yueming and Li, Rui and Cao, Xuan and Gao, Leon and Gong, Zhaojie and Gu, Fangda and He, Jiayuan and Lu, Yinghai and Shi, Yu},
    title = {Actions Speak Louder than Words: Trillion-Parameter Sequential Transducers for Generative Recommendations},
    booktitle = {Proceedings of the 41st International Conference on Machine Learning},
    series = {Proceedings of Machine Learning Research},
    volume = {235},
    pages = {58484--58509},
    year = {2024},
    publisher = {PMLR},
    url = {https://proceedings.mlr.press/v235/zhai24a.html},
}

@inproceedings{park2025laus,
    author = {Park, Choongwon},
    title = {{LLM} as User Simulator: Towards Training News Recommender without Real User Interactions},
    booktitle = {Proceedings of the 48th International ACM SIGIR Conference on Research and Development in Information Retrieval},
    series = {SIGIR '25},
    pages = {3080--3084},
    year = {2025},
    address = {New York, NY, USA},
    publisher = {Association for Computing Machinery},
    doi = {10.1145/3726302.3730224},
    url = {https://doi.org/10.1145/3726302.3730224},
}

@inproceedings{allouah2026agentic,
    author = {Allouah, Amine and Besbes, Omar and Figueroa, Josu{\'e} D. and Kanoria, Yash and Kumar, Akshit},
    title = {What Is Your {AI} Agent Buying? Evaluation, Biases, Model Dependence, \& Emerging Implications of Agentic E-Commerce},
    booktitle = {Proceedings of the ACM Web Conference 2026},
    series = {WWW '26},
    pages = {8697--8700},
    year = {2026},
    address = {New York, NY, USA},
    publisher = {Association for Computing Machinery},
    doi = {10.1145/3774904.3792943},
    url = {https://doi.org/10.1145/3774904.3792943},
}

@article{li2025tridecoupled,
    author = {Li, Xiaopeng and Chen, Bo and She, Junda and Cao, Shiteng and Wang, You and Jia, Qinlin and He, Haiying and Zhou, Zheli and Liu, Zhao and Liu, Ji and Zhang, Zhiyang and Zhou, Yu and Tang, Guoping and Yang, Yiqing and Guo, Chengcheng and Dong, Si and Cai, Kuo and Jia, Pengyue and Wang, Maolin and Wang, Wanyu and Wang, Shiyao and Luo, Xinchen and Hu, Qigen and Luo, Qiang and Lv, Xiao and Ma, Chaoyi and Tang, Ruiming and Gai, Kun and Zhou, Guorui and Zhao, Xiangyu},
    title = {A Survey of Generative Recommendation from a Tri-Decoupled Perspective: Tokenization, Architecture, and Optimization},
    journal = {Preprints},
    month = dec,
    year = {2025},
    doi = {10.20944/preprints202512.0203.v1},
    url = {https://doi.org/10.20944/preprints202512.0203.v1},
}

\end{document}